\documentstyle[preprint,eqsecnum,aps,prb,amstex]{revtex}
\begin{document}
\title{Non-linear electrical response in a charge/orbital ordered $\Pr_{0.63}$Ca$%
_{0.37}$MnO$_3$ crystal : the charge density wave analogy.}
\author{A.\ Wahl, S.\ Mercone, A.\ Pautrat, M.\ Pollet, Ch. Simon, D.\ Sedmidubsky$%
^{*}$}
\address{Laboratoire CRISMAT, UMR 6508, Institut desSciences de la Mati\`{e}re et du\\
Rayonnement - Universit\'{e} de Caen, 6 Boulevard du Mar\'{e}chal Juin,\\
14050 Caen Cedex, France.\\
$^{*}$ Institute of Physics of ASCR, Cukrovarnicka 10. 16200 Prague 6, Czech%
\\
Republic}
\date{\today}
\maketitle

\begin{abstract}
Non-linear conduction in a charge-ordered manganese oxide Pr$_{0.63}$Ca$%
_{0.37}$MnO$_3$ is reported. To interpret such a feature, it is usually
proposed that a breakdown of the charge or orbitally ordered state is
induced by the current. The system behaves in such a way that the bias
current may generate metallic paths giving rise to resistivity drop. One can
describe this feature by considering the coexistence of localized and
delocalized electron states with independent paths of conduction. This
situation is reminiscent of what occurs in charge density wave systems where
a similar non-linear conduction is also observed. In the light of recent
experimental results suggesting the development of charge density waves in
charge and orbitally ordered manganese oxides, a phenomenological model for
charge density waves motion is used to describe the non-linear conduction in
Pr$_{0.63}$Ca$_{0.37}$MnO$_3$. In such a framework, the non-linear
conduction arises from the motion of the charge density waves condensate
which carries a net electrical current.
\end{abstract}

\pacs{}

Charge ordering (CO ) is a common characteristic of transition metal oxides
with perovskite structure.\cite{CO} In colossal magnetoresistive (CMR)
manganites, the formation of an $Mn^{3+}:Mn^{4+}$ ordered phase is the most
obvious manifestation of the $e_g$ electrons localization.\cite{REVIEW,CMR}
The physics of the phase transition from a CO - antiferromagnet to
charge-delocalized ferromagnet have been extensively studied in recent
years. This CO destabilization is of great interest because this feature can
be achieved under a wide variety of external perturbations.\cite
{LEE95,TOM96,YOS95,YOS96,TOM95,ANA99} For instance, numerous experimental
results have shown that the application of a moderate electric field leads
to an insulator - metal (I-M) transition associated with a strong
non-linearity of the voltage - current (V-I) characteristics.\cite
{ASA97,STA00,GUH00a,GUH00b,RAO00,BUD01} To interpret such a phenomenon it is
proposed that a breakdown of the CO state is induced by the current and that
metallic filaments are created.\ However, when considering the properties of
manganites with colossal magnetoresistance, one have also to take into
account, besides the charge and spin degrees of freedom, the orbital
structure of the transition metal ions. Therefore, the description of the
collective behavior of $e_g$ carriers in connection with the specific kind
of charge $and$ orbital ordering (OO) (which depend on the doping state)
might be of fundamental interest for the understanding of the non-linear
electrical response of the CO / OO manganites.\cite
{ASA97,STA00,GUH00a,GUH00b,RAO00,BUD01}

Compounds of the system Pr$_{1-x}$Ca$_x$MnO$_3$ with $x$ close to 0.4 is a
well documented material for which extensive investigations have been
carried out by magnetic and transport measurements.\cite
{LEE95,TOM96,SMO00,LEE99,MAR99,MAI97,HAR01}\ Recently, in Pr$_{0.63}$Ca$%
_{0.37}$MnO$_3,$Asaka et al.\cite{ASAKA02}, by means of low-temperature
electron microscopy have revealed new forms of structural modulations that
are discussed in terms of temperature-dependent ordering of $e_g$ electrons.
These authors have observed superlattice reflection spots with a modulation
wave vector $q_1=(0,\frac 12,0)$ below 230 K, suggesting the classical
formation of $d_{3x^2-r^2}/d_{3y^2-r^2\text{ }}$orbital ordering, similar to
the half-doped manganite Pr$_{0.5}$Ca$_{0.5}$MnO$_3.$ Below 150 K, a new
modulation wave vector appears, $q_2=(\frac 14,\frac 14,\frac 12).\;$In this
temperature range, $Mn^{3+}$ ions must be substituted partially on the $%
Mn^{4+}$ sublattice in the $x=\frac 12$- type CO / OO in the ratio 1:3 .
This kind of ordering can be viewed as a quasi-one dimensional electronic
structure with a reduced dimensionality compared to the 1:1 CO / OO of Pr$%
_{0.5}$Ca$_{0.5}$MnO$_3$. This low temperature electronic structure is
described as follows along the $c-axis$ : a $d_{3x^2-r^2}/d_{3y^2-r^2\text{ }%
}$orbital ordering of $Mn^{3+}$ occurs while in an adjacent plane, the $%
d_{3z^2-r^2}$ orbitals of the $Mn^{3+}$ are separated by three $Mn^{4+}$
ions. Such a periodic arrangement (non commensurate) of charge in Pr$_{0.63}$%
Ca$_{0.37}$MnO$_3$ might lead to the development of a charge density wave
(CDW) condensate.\ Such a CDW formation has also been deduced recently in
manganese oxides of this composition by means of angle resolved
photoemission spectroscopy. \cite{CHU01} By means of THz time-domain
spectroscopy,\cite{KIDA02} evidence for a charge density wave condensate in
the CO / OO manganite Pr$_{0.7}$Ca$_{0.3}$MnO$_3$ has been reported. For
this composition, Kida et al.\cite{KIDA02} have revealed the existence of a
finite peak structure well below the charge gap. They attributed this
observed structure to the collective excitation mode arising from a charge
density wave condensate.

The more fascinating properties of CDW systems are related to nonlinear
conduction induced by an electric field.\cite{GRUNER} This feature was
interpreted as the conductivity associated with the motion of CDW. As
proposed in\cite{LEERICE} , the phase of the CDW could be pinned by several
mechanisms ; if an electric field strong enough to overcome the pinning
energy were applied, the CDW can be depinned and carries a current. In the
following, on the basis of the experimental observations suggesting a CDW
development in manganese oxides\cite{KIDA02,CHU01,CALVA}$,$we tentatively
account for non-linear V-I characteristics in Pr$_{0.63}$Ca$_{0.37}$MnO$_3$
by considering a phenomenological model for CDW motion.\cite{MONCEAU}

Using the floating-zone method with a feeding rod of nominal composition Pr$%
_{0.6}$Ca$_{0.4}$MnO$_3$, a several-cm-long single crystal was grown in a
mirror furnace. X-ray diffraction and electron diffraction studies, which
were performed on pieces coming from the same part of the crystal, attested
that the samples are single phased, and well crystallized. The cell is
orthorhombic with a Pnma space group, in agreement with previously reported
structural data. The electron diffraction (ED) investigations have shown the
existence of twinning domains, which result from the reversible phase
transitions (from cubic to orthorhombic) undergone by the crystals upon
cooling. The cationic composition derived from energy-dispersive x-ray
analyses leads to the formula Pr$_{0.63}$Ca$_{0.37}$MnO$_3$. Four linear
contact pads of In were soldered onto the sample in linear four-probe
configuration.\ $V-I$ data were taken with current biasing (Keithley 236)
and with a temperature control of $100$ $mK$.\ 

In figure 1, the temperature variation of the resistivity $(\rho $ $vs$ $T)$
of a Pr$_{0.63}$Ca$_{0.37}$MnO$_3$ crystal is shown for various bias current
under zero magnetic field. For low current, the $\rho $ $vs$ $T$ curve
displays a pronounced semiconducting-like behavior, leading to an insulating
state at low temperatures. \ Our current source overloads for $V_{\lim
it}=100V$, thus the maximum measurable resistance for $10^{-3}mA$ is around $%
100M\Omega $.\ For higher current, the insulating state is no longer
observable ; the resistivity is indeed strongly depressed. There is no true
insulator - metal transition but a trend to saturation when the temperature
is lowered. Moreover, for high and low currents, one can also denote a clear
kink around 230K, which related to the charge ordering phenomenon. Figure 2
shows V-I characteristics for various temperatures (60K, 80K, 120K and
140K). A strong non-linearity i. e. a deviation from the Ohm's law, is
observed when the bias current attains a threshold value (0.66 A.cm$^{-2}$
at 80K).\ This non-linearity is even more obvious when $\frac R{Rohmic}$ $vs$
$I$ curves are plotted (See figure 3). As expected the resistance is
independent of the bias current in the ohmic regime and is strongly
decreased for a critical value of the current. One can observe that the
current value where the non-linearity sets in and the width of the
transition are strongly temperature dependent (both increase as the
temperature increases). Because of the rounding of the variation of $\frac R{%
Rohmic}$ $vs$ $I$, it is very difficult to define a value for a critical
current.

We have carefully checked that the Joule heating is irrelevant to account
for this current induced effect.\ The temperature rise of the sample with
respect to the sample holder $\left( \Delta T\right) $ has been measured by
attaching a thermometer on the top of the sample itself. In the low
temperature range and for the highest power dissipation, one measure $\Delta
T\prec 6K$. For higher temperature of measurements, $\Delta T$ becomes
negligible. Moreover, a great amount of papers have verified this point in
samples of the same composition. The peculiar electrical response that we
describe here is in remarkable agreement with previous studies carried out
on CO / OO manganese oxides.\cite{ASA97,STA00,GUH00a,GUH00b,RAO00,BUD01} To
account for the nonlinear conduction in CO / OO manganese oxides, a special
type of dielectric breakdown of the CO state brought about by small electric
field is invoked. It is argued that the barrier between the CO state and the
metallic state is small enough that a small electric field can cause an I-M
transition.\cite{ASA97,STA00,GUH00a,GUH00b,RAO00,BUD01}\ The system behaves
in such a way that the bias current may generate metallic path giving rise
to resistivity drop. One can describe this feature by considering
coexistence of localized and delocalized electron states with independent
path of conduction. This situation is analogous to what occurs in charge
density waves systems.\cite{GRUNER}

In the following, we propose that, for Pr$_{0.63}$Ca$_{0.37}$MnO$_3,$ the
non-linearity observed in V-I characteristics can be linked to the
collective excitation of the CDW condensate. By analogy with the sliding
mode conductivity in NbSe$_3$,\cite{GRUNER,MONCEAU} the non-linear transport
properties are explained by the motion of charge density waves, which carry
a net electrical current, superimposed to the poorly mobile carriers of the
system, when the current reaches a critical current necessary to overcome
the pinning forces acting on the CDW condensate. The total current density
can then be written : $j=j_{CDW}+\sigma E$ where $j_{CDW}$ is the current
density associated with the motion of the CDW condensate and $\sigma $ is
the ohmic conductivity at low electric field / current. The dynamic of the
collective mode is described in terms of a position- and time-dependent
order parameter, thus both amplitude and phase fluctuation occur. The
relevant collective mode is the phase one which is gapless. As described by
Gr\"{u}ner et al.\cite{GRUNER}, the phase mode carries a dipole moment as it
corresponds to the motion of condensed electrons through the background of
position changes of the ions ; consequently, the phase mode carries a
current for a wave vector $q=0$. Monceau et al.\cite{MONCEAU} have developed
a model where the phase of the CDW is described as an overdamped oscillator.
The non-linear electrical response is the consequence of the interaction
between the phase mode and the lattice. These interactions lead to a finite
pinning energy and consequently to a finite bias current for the non-linear
conduction to set in. In the manganese oxides of the system Pr$_{1-x}$Ca$_x$%
MnO$_3$, the pinning centers that we are dealing with are not well
identified. The pinning is expected on general grounds, e. g. the L/A
substitutional disorder that acts as a quenched disorder, other lattice
defects or possible homogeneities. To be more precise, in these materials,
the Jahn-Teller ions Mn3+ has a major role. The Jahn-Teller induced
distortion of the Mn3+O6 octahedra can act as a periodic pinning site due to
local strain field. Very early in the subject\cite{HER96}, these strain
fields have been observed by high resolution electron microscopy in Pr
manganites. Existence of monoclinic domains in the orthorhombic matrix
corresponding to a modification of the geometry of the Mn$^{3+}$O$_6$
octahedra has been reported. At the junction between the two kinds of
domains, a defective structure is observed and can be interpreted as a twin
boundary. In the Pr managanites, another very important feature deals with
the existence of a rather large density of isolated point defects. Once
again, this can be associated with the local modification of the manganese
environment owing to the Jahn-Teller effect.

\smallskip Neglecting inertial effects and considering the phase $\phi (r,t)$
as uniform through out a given domain, the equation of motion of the phase
obtained in the framework of the ''single particle'' model is :

\[
\tau \frac{\partial \phi }{\partial t}(1+\beta )=\frac j{\sigma E_c}+\sin
\phi 
\]

where $\phi $ is the phase of the domain, $\beta $ is an experimental
parameter defined as $\beta =\frac{\sigma _{E\rightarrow \infty }-\sigma
_{E\rightarrow 0}}{\sigma _{E\rightarrow 0}}$, $\tau $ is a dimensionless
parameter. When the applied current is constant the average electric field
is :

\[
E=\frac j\sigma -\frac{\beta E_c}{1+\beta }\left[ \left( \frac j{j_c}\right)
^2-1\right] ^{\frac 12} 
\]

where $\sigma $ is the ohmic conductivity, $E_c$ and $j_c$ are the critical
electric field and current, respectively, where the non linearity sets in.
Introducing the experimental variables $R$, the measured resistance and $I$,
the applied current, one obtains :

\[
R=R_{ohmic}\left[ 1-\left( \frac \beta {1+\beta }\right) \left[ 1-\frac{I_c^2%
}{I^2}\right] ^{\frac 12}\right] 
\]

According to Monceau et al.\cite{MONCEAU}, to account for a more realistic
description of the sample, we assume that it is formed of multiple domains.
A statistical distribution of domains is then considered, each one being
depinned for a given value of $I_c$, in a mean field approximation. Let us
call $P(I_c)dI_c$, the normalized probability to observe the depinning of a
domain for a current value in the range $I_c+dI_c$ and $I_c$. $m$ is the
mean value of this statistical distribution and $s$, the standard deviation.
The resistance of the system is then given by :

\[
R=R_{ohmic}\left[ 1-\left( \frac \beta {1+\beta }\right)
\int_{I_{Ohmic}}^IP(I_c)\left[ 1-\frac{I_c^2}{I^2}\right] ^{\frac 12%
}dI_c\right] 
\]

$I_{Ohmic}$ is a cutoff corresponding to the current for which the first
domain is depinned. The range $I<I_{Ohmic}$ defines the true ohmic response.

This model is essentially phenomenological. The main assumption of the mean
field approach is to assume a statistical distribution for critical
currents. Originally, Monceau et $al$.\cite{MONCEAU} have used a gaussian
distribution without any justification. According to the literature, the
properties of the CDW systems are very similar to the typical properties of
the wide range of glass materials. The description of the behaviour of many
glass materials often uses the hypothesis about dynamic scaling. On this
base, the dynamic effects in disordered glass material have been considered.%
\cite{SOUL91,SOUL94,OGI85,CAVA84} In particular, it was shown that in
correlated systems with some disorder, the lognormal distribution for
relaxation times is relevant.\cite{SOUL91,SOUL94,OGI85,CAVA84} The
expression of the distribution we have used is : \smallskip 
\[
P(I_c)=\frac 1{I_c\beta \sqrt{2\pi }}\exp -[\frac{\left( \ln I_c-\ln \alpha
\right) ^2}{2\beta ^2}] 
\]
with $\alpha =m\sqrt{k}$, $\beta =\ln k$ and $k=\left( \frac sm\right) ^2+1$%
. $m$ and $s$ are the mean and the standard deviation respectively.

The results are shown in figure 4 for T = 80K. One observe a good agreement
between the predicted model (solid lines) and the experimental $R$ $vs$ $I$
curves (symbols). The same agreement is found for other temperatures. Figure
4 can not be considered as experimental evidence by itself. Nevertheless, in
the first part of the paper, we discuss the experimental evidences for the
occurrence of a charge density wave in the CO manganites. Within this
framework, the issue is then to test if a model, which has been successfully
applied for archetypal CDW systems, can describe our data. Thus, figure 4
suggests that an analogy between the CDW domains sliding and the non-linear
conduction in CO manganites might be reliable. In figure 5, the
distributions calculated from the fit parameters are shown for various
temperatures . In addition, the temperature dependence of the standard
deviation ($s$) and mean ($m$) are reported in figure 6. Those two
parameters are of great interest since their temperature dependence can be
understood in light of our knowledge concerning the stability of the CO / OO
state in Pr$_{0.63}$Ca$_{0.37}$MnO$_3$.\cite{ASAKA02} The temperature
dependence of $s$ (main panel figure 6) gives a good picture of the
homogeneity of the CO / OO state (that is at the origin of the development
of the CDW condensate). From low temperature up to 100 K, there is only a
slight variation of $s$ with the temperature while above this temperature,
the value of $s$ is substantially increased. This feature can be related
with the two modulation structure observed by Asaka et al.\cite{ASAKA02}
Below 230 K, a CO / OO state of the $d_{3x^2-r^2}/d_{3y^2-r^2\text{ }}$%
orbitals of the $Mn^{3+}$ similar to the half-doped Pr$_{0.5}$Ca$_{0.5}$MnO$%
_3$ remains essential ((CO / OO)$_{q1})$. While the temperature is
decreased, a new modulation structure corresponding to a partial
substitution of $Mn^{3+}$ on the $Mn^{4+}$ sublattice in ratio 1:3 appears
((CO / OO)$_{q2})$. The (CO / OO)$_{q2}$, which is long range for $T=100K$,
modifies the uniformity of the charge density, leading to the development of
a CDW condensate. Our $s$ $vs$ $T$ data can be understood on the basis of
this observation. At low temperature, the (CO / OO)$_{q2}$ state is rather
homogenous and long range, leading to a small width of the statistical
distribution of $I_c$ since there is not a great variety of domains that
have to be depinned. As the temperature increases, the (CO / OO)$_{q2}$ is
no longer long range and a fragmentation of the CDW lattice into smaller
domains occurs. Consequently, one observes an increase of the width of the
distribution. The temperature dependence of the mean is less easy to
understand (inset figure 6). It appears that the current necessary to depin
a domain increases with temperature. Arguing from analogy with the NbSe$_3$
system, this feature might be related to the size of the (CO / OO)$_{q2}$
domains where the CDW is developed. In the model proposed by Monceau et al.%
\cite{MONCEAU}, the interactions between pinning centers and the phase of
the CDW can be written : $W_{interact}=-V_0\cos \left[ \phi (r_i)-\psi
(r_i)\right] $ where $r_i$ is the position of a pinning center, $\phi $, the
phase of the CDW and $\psi $, the ideal value of the phase for an effective
pinning. Below 230 K, there is a nucleation of small (CO / OO)$_{q2}$ / CDW
domains at places where the random distribution of pinning centers (i. e of $%
\psi (r_i)$ values) fixes the phase of the domain in an attractive position
corresponding to higher depinning current. As the temperature decreases down
to 100 K, the domains are growing, encompassing less favorable value of $%
\psi $ and consequently, lower currents are necessary to depin them.

In conclusion, V-I characteristics have been measured in a Pr$_{0.63}$Ca$%
_{0.37}$MnO$_3$ crystal. Strong non-linear conduction is observed under zero
field for a wide range of temperature. This feature is reminiscent of what
is observed in charge density waves systems. On the basis of the
experimental observations suggesting a CDW development in manganese oxides$,$%
we speculate that non linear conduction in Pr$_{0.63}$Ca$_{0.37}$MnO$_3$ can
be described by considering a phenomenological model for CDW motion.

\section{Figures Captions}

\begin{description}
\item[Figure 1]  : Temperature dependence of resistance for a $\Pr_{0.63}$Ca$%
_{0.37}$MnO$_3$crystal with various bias currents.

\item[Figure 2]  : $V-I$ characteristics under zero field for various
temperatures ($60K,$ $80K,120K$ and $140K).$

\item[Figure 3]  : $\frac R{R_{Ohmic}}$ versus bias current for various
temperatures.

\item[Figure 4]  : $R$ versus bias current at 80K. The solid line is the fit
using the model described in the text.

\item[Figure 5]  Shape of the distributions of $I_c$ for various temperatures

\item[Figure 6]  Temperature dependence of the standard deviation (s) of the
distribution. Inset : Temperature dependence of the mean (m) of the
distribution
\end{description}

\end{document}